
\input phyzzx
\hoffset=1truein
\voffset=1.0truein
\hsize=6truein
\def\TITLEPAGE{\frontpagetrue}
\def\CALT#1{\hbox to\hsize{\tenpoint \baselineskip=12pt
	\hfil\vtop{\hbox{\strut CALT-68-#1}
	\hbox{\strut DOE RESEARCH AND}
	\hbox{\strut DEVELOPMENT REPORT}}}}

\def\CALTECH{\smallskip
	\address{California Institute of Technology, Pasadena, CA 91125}}

\def\AUTHOR#1{\vskip .5in \centerline{#1}}

\def\ABSTRACT#1{\vskip .5in \vfil \centerline{\twelvepoint \bf Abstract}
	#1 \vfil}
\def\ENDTITLEPAGE{\vfil\eject\pageno=1}

\def\sqr#1#2{{\vcenter{\hrule height.#2pt
      \hbox{\vrule width.#2pt height#1pt \kern#1pt
        \vrule width.#2pt}
      \hrule height.#2pt}}}

\def\section#1#2{
\noindent\hbox{\hbox{\bf #1}\hskip 10pt\vtop{\hsize=5in
\baselineskip=12pt \noindent \bf #2 \hfil}\hfil}
\medskip}

\def\underwig#1{	
	\setbox0=\hbox{\rm \strut}
	\hbox to 0pt{$#1$\hss} \lower \ht0 \hbox{\rm \char'176}}

\def\bunderwig#1{	
	\setbox0=\hbox{\rm \strut}
	\hbox to 1.5pt{$#1$\hss} \lower 12.8pt
	 \hbox{\seventeenrm \char'176}\hbox to 2pt{\hfil}}

\def\MEMO#1#2#3#4#5{
\frontpagetrue
\centerline{\tencp INTEROFFICE MEMORANDUM}
\smallskip
\centerline{\bf CALIFORNIA INSTITUTE OF TECHNOLOGY}
\bigskip
\vtop{\tenpoint
\hbox to\hsize{\strut \hbox to .75in{\caps to:\hfil}\hbox to 3.8in{#1\hfil}
\quad\the\date\hfil}
\hbox to\hsize{\strut \hbox to.75in{\caps from:\hfil}\hbox to 3.5in{#2\hfil}
\hbox{{\caps ext-}#3\qquad{\caps m.c.\quad}#4}\hfil}
\hbox{\hbox to.75in{\caps subject:\hfil}\vtop{\parindent=0pt
\hsize=3.5in #5\hfil}}
\hbox{\strut\hfil}}}

\tolerance=10000
\hfuzz=5pt

\TITLEPAGE
\CALT{1850}         
\bigskip
\titlestyle {Black Holes with a Generalized Gravitational Action
\foot{Work supported in part by the U.S. Dept. of Energy
under Contract no. DEAC-03-81ER40050.}}
\AUTHOR{Ming Lu and Mark B. Wise}
\CALTECH
\ABSTRACT{Microscopic black holes are sensitive to higher dimension
operators in the gravitational action.  We compute the influence of
these operators on the Schwarzschild solution using perturbation
theory.  All (time reversal invariant) operators of dimension six are
included (dimension four operators don't alter the Schwarzschild
solution).  Corrections to the relation between the Hawking temperature
and the black hole mass are found.  The entropy is calculated using the
Gibbons-Hawking prescription for the Euclidean path integral and using
naive thermodynamic reasoning.  These two methods agree, however, the
entropy is not equal to 1/4 the area of the horizon.}
\ENDTITLEPAGE

\eject

Classical black hole solutions to generalizations of Einsteins gravitational
action
$$	S_0 = {1\over 16\pi G} \int d^4 x \sqrt{g} R \,\, , \eqno (1)$$
have been found in several different cases.  Particularly noteworthy is
the work of Callan, Myers and Perry and Myers who (motivated by string theory)
examined black holes in a model with a dilaton field and terms quadratic
and quartic in the curvature tensor.$^{[1]}$  Our methods and
conclusions are similar to those in Ref. [1].  The main differences are
that we consider pure gravity (i.e., no dilaton field) and allow
arbitrary coefficients for the higher powers of the curvature tensor
instead of the particular combinations implied by string theory.  In this
paper we use perturbation theory to
find the corrections to the Schwarzschild solution that arise from
including higher powers of the curvature tensor in the gravitational
action.  All possible time reversal invariant terms of dimension four and
dimension
six are included, although only a few of the dimension six operators
actually influence the solution.  Such terms are expected to occur, for
example, from integrating out massive degrees of freedom.  Their
influence on macroscopic black holes is negligible but they may play a
role in the physics of microscopic black holes in the final stages of
the evaporation process.$^{[2,3]}$

In general the gravitational action can be written as$^{[4]}$
$$	S = S_0 + S_1 + S_2 + ... \eqno (2)$$
where $S_i$ denotes the contribution of operators of dimension $2i + 2$
($S_0$ is given by eq. (1)).  The spacetime metric generated by any
static spherically symmetric matter distribution can be cast in the
form$^{[5]}$
$$	ds^2 = - e^{-2\phi (r)} [1 - b(r)/r]dt^2 + {dr^2\over [1 - b(r)/r]}$$
$$	+ r^2 (d\theta^2 + \sin^2 \theta d\phi^2) \,\, . \eqno (3)$$
Outside the matter distribution the metric satisfies the equations of
motion
$$	G^{\mu\nu} =  8\pi G \sum_{i = 1}^\infty {\cal T}_i^{\mu\nu} \,\, ,
\eqno (4)$$
with
$$	{\cal T}_i^{\mu\nu} = {2\over\sqrt{g}} {\delta S_i\over \delta g_{\mu\nu}}
\,\, .
 \eqno (5)$$

The Schwarzschild black hole solution corresponds to ${\cal T}_i^{\mu\nu} =
0$ for $i = 1,2,...$ and
has $b = 2GM$ and $\phi = 0$, where $M$ is the black hole mass.  We are
interested in corrections to this that arise from the $S_i, i = 1,2,...$
.  In fact the contribution of operators of dimension four to the
gravitational action, $S_1$, does not alter this solution.  The most
general (time reversal invariant) form for $S_1$ is
$$	S_1 = \int d^4 x \sqrt{g} \{a_1 R^2 + a_2 R_{\mu\nu}
R^{\mu\nu}\} \,\, . \eqno (6)$$
No term of the form $R_{\alpha\beta\gamma\delta}
R^{\alpha\beta\gamma\delta}$ needs to be considered since
$R_{\alpha\beta\gamma\delta} R^{\alpha\beta\gamma\delta} - 4 R_{\mu\nu}
R^{\mu\nu} + R^2$ is a total derivative.$^{[6]}$  $S_1$ does not alter the
Schwarzschild solution because eq. (6) implies that the terms in
${\cal T}_1^{\mu\nu}$ contain a factor of $R$ or $R^{\mu\nu}$ which vanish for
the Schwarzschild solution.

In this paper we compute the corrections to the metric that arise from
$S_2$.  The most general (time reversal invariant) form for $S_2$ is
$$	S_2 = \int d^4 x \sqrt{g} \{c_1 R_{\mu\nu\lambda\sigma}
R^{\alpha\beta\lambda\sigma} R^{\mu\nu}_{~~~\alpha\beta} + c_2
R^{\mu\nu}_{~~~\lambda\sigma} R_{\mu\alpha}^{~~~~\lambda\beta}
R^{\alpha\sigma}_{~~~\nu\beta}$$
$$	+ c_3 R_{\mu\nu} R^{\mu\alpha\beta\gamma} R^\nu_{~~\alpha\beta\gamma}
+ c_4 R R_{\mu\nu\lambda \kappa} R^{\mu\nu\lambda \kappa}$$
$$	+ c_5 R_{\mu\nu} R_{\lambda \kappa} R^{\mu\lambda\nu \kappa} + c_6
R_{\mu\nu} R^{\mu\lambda} R_\lambda^{~~\nu} + c_7 R^3$$
$$	+ c_8 R R_{\mu\nu} R^{\mu\nu}\} \,\, . \eqno (7)$$
The identity
$$	R_{\alpha\beta\gamma\delta; \mu}^{~~~~~~~~;\mu} =
R_{\alpha\beta\rho\sigma} R_{\gamma\delta}^{~~~\rho\sigma} + 2
(R_{\alpha\rho\gamma\sigma} R_{\beta~~~\delta}^{~~~\rho~~\sigma} -
R_{\alpha\rho\delta\sigma} R_{\beta~~~\gamma}^{~~~\rho~~\sigma}) \eqno (8)$$
has been used to express terms containing two curvature tensors (i.e.,
$R, R_{\mu\nu}$ or $R_{\mu\nu\alpha\beta}$) and two covariant
derivatives in terms of those in eq. (7).  For example, using the
Bianchi identity it can be shown that
$$	R_{\alpha\beta\gamma\delta; \mu} R^{\alpha\beta\gamma\mu
;\delta} = - {1\over 2} (R_{\alpha\beta\gamma\delta}
R^{\alpha\beta\delta\gamma;\mu})_{;\mu} + {1\over 2}
R_{\alpha\beta\gamma\delta} R_{~~~~~~~~;\mu}^{\alpha\beta\delta\gamma;\mu} \,\,
{}.
\eqno (9)$$
The symmetries of the Riemann tensor imply that the terms proportional
to $c_1$ and $c_2$ are the most general ones containing three four-index
Riemann tensors.  For example it is easy to see that
$$	R_{\mu\nu\lambda\sigma} R^{\mu\nu\alpha\beta}
R_{\alpha~~\beta}^{~~\lambda~~~\sigma} = {1\over 2} R_{\mu\nu\lambda\sigma}
R^{\alpha\beta\mu\nu} R_{\alpha\beta}^{~~~\lambda\sigma}\,\, , \eqno
(10a)$$
and
$$	R^{\mu\nu}_{~~~\lambda\sigma} R_{\mu\alpha}^{~~~\lambda\beta}
R_{\nu~~~~\beta}^{~~\sigma\alpha} = {1\over 2} R_{\mu\nu\lambda\sigma}
R^{\mu\nu\alpha\beta} R_{\alpha~~\beta}^{~~\lambda~~~\sigma} \,\, . \eqno
(10b)$$

Using the same reasoning as for $S_1$ it can be shown that the terms
proportional to $c_5, c_6, c_7$ and $c_8$ in eq. (7) do not alter the
Schwarzschild solution.  The effective stress energy tensor ${\cal T}_2$ that
arises from $S_2$ is
$$	{\cal T}_2^{~\eta\phi} = c_1 \bigg[g^{\eta\phi}
R_{\mu\nu\lambda\sigma} R^{\alpha\beta\lambda\sigma}
R^{\mu\nu}_{~~~\alpha\beta} - 6 R^\eta_{~~\nu\lambda\sigma}
R^{\phi\nu}_{~~~\alpha\beta} R^{\alpha\beta\lambda\sigma}$$
$$	+ 6 (R^{\eta\nu}_{~~~\kappa\ell} R^{\kappa\ell\lambda\phi})_{;\lambda
;\nu} + 6 (R^{\phi\nu}_{~~~\kappa\ell} R^{\kappa\ell\lambda\eta})_{;\lambda
;\nu}\bigg]$$
$$	c_2 \bigg[g^{\eta\phi} R^{\mu\nu}_{~~~\lambda\sigma}
R_{\mu\alpha}^{~~~\lambda\beta} R^{\alpha\sigma}_{~~~\nu\beta} -
6 R^{\eta\nu}_{~~~\lambda\sigma} R^{\phi\alpha\lambda\beta}
R_{\alpha~~\nu\beta}^{~~\sigma}$$
$$	- 3 (R^{\eta\rho\phi\delta}
R_{\rho~~~\delta}^{~~\kappa\nu})_{; \kappa
;\nu} - 3 (R^{\phi\rho\eta\delta}
R_{\rho~~~~\delta}^{~~\kappa\nu})_{; \kappa
;\nu}$$
$$	+ 3 (R^{\eta\rho \kappa\delta} R_{\rho~~~\delta}^{~~\phi\nu})_{;
\kappa ;\nu} + 3 (R^{\phi\rho \kappa\delta}
R_{\rho~~~\delta}^{~~\eta\nu})_{; \kappa
; \nu}\bigg]$$
$$	+ c_3 \bigg[-
(R^{\alpha\beta\gamma\eta}R_{\alpha\beta\gamma}^{~~~~~\phi})^{;\lambda}_
{~~;\lambda} -  g^{\eta\phi}
(R^{\alpha\beta\gamma\lambda}
R_{\alpha\beta\gamma}^{~~~~~\sigma})_{;\sigma;\lambda}$$
$$+ (R_{\alpha\beta\gamma}^{~~~~\eta}
R^{\alpha\beta\gamma\kappa})^{;\phi}_{~~~;\kappa} +
(R_{\alpha\beta\gamma}^{~~~~\phi}
R^{\alpha\beta\gamma\kappa})^{;\eta}_{~~~;\kappa} \bigg]$$
$$	+  c_4 \bigg[- 2 g^{\phi\eta} (R_{\mu\nu\alpha\beta}
R^{\mu\nu\alpha\beta})_{~~~;\lambda}^{; \lambda} + (R_{\mu\nu\alpha\beta}
R^{\mu\nu\alpha\beta})^{;\phi ;\eta} + (R_{\mu\nu\alpha\beta}
R^{\mu\nu\alpha\beta})^{;\eta;\phi}\bigg]$$
$$	+ ... \eqno (11)$$
The ellipsis in eq. (11) denotes terms involving the Ricci tensor and
the curvature scalar.  We compute the influence of $c_1,c_2,c_3$ and $c_4$ on
the metric using perturbation theory about the Schwarzschild
black hole solution.  In the lowest nontrivial order of perturbation
theory the equation for the metric $G^{\mu\nu} =  8 \pi G{\cal T}_2^{\mu\nu}$
has the effective stress energy tensor ${\cal T}_2^{\mu\nu}$ evaluated in the
Schwarzschild background, and the Einstein tensor $G^{\mu\nu}$
evaluated at linear order$^{[7]}$ in the perturbation about this
metric.  The $\hat r\hat r$ and $\hat t\hat t$ components of this
equation yield the following differential equations for $b$ and $\phi$.
$$	b'(r) = 48\pi G \bigg[(-98c_1 - 25c_2 - 33c_3 - 132c_4)
(2GM)^3/r^7 $$
$$+ (90c_1 + {45\over 2} c_2 + 30c_3 + 120c_4) (2GM)^2/r^6\bigg] \eqno
(12a)$$
$$	\phi'(r) = 48\pi G (- 54c_1 - 18c_2 - 21c_3 - 84c_4) (2GM)^2/r^7\,\, .
\eqno (12b)$$
The function $b(r)$ and $\phi(r)$ are determined by integrating eqs.
(12) with the boundary conditions $b(\infty) = 2GM$ and $\phi(\infty) =
0$.  The horizon is located at a radius $r_H$ that satisfies $b(r_H) =
r_H$.  The relationship between the radius of the horizon and the black
hole mass is
$$	r_H = 2GM \left[1 - {\pi\over G^3 M^4} \left(5c_1 + c_2 +
{3\over 2} c_3 + 6c_4\right)\right] \,\, . \eqno (13)$$

The Euclidean section of the geometry (3) has ${\cal R}^2 \times {\cal S}^2$
topology.  The imaginary time coordinate plays the role of the polar
angle in ${\cal R}^2$ and the origin of this plane is at the horizon.
The period of the imaginary time coordinate is the inverse of the
Hawking temperature $T$.  Expanding the metric (3) in the vicinity of
the horizon gives$^{[5]}$
$$	T = {1\over 4\pi r_H} e^{-\phi(r_H)} (1 - b'(r_H)) \,\, . \eqno
(14)$$
Using this and our results for $b$ and $\phi$ we find that when the
terms in $S_2$ are included the relationship between the Hawking
temperature and mass becomes
$$	T = {1\over 8\pi GM} \left[ 1 + {\pi\over G^3 M^4} \left(2 c_1 -
{1\over 2} c_2\right)\right]\,\, . \eqno (15)$$

Note that $c_3$ and $c_4$ do not affect the relationship between the
Hawking temperature and the mass of the black hole.  The terms
proportional to $c_3$ and $c_4$ in the action $S_2$ can be removed by a
field redefinition of the metric.  For example, the redefinition
$$	g_{\mu\nu} \rightarrow g_{\mu\nu} (1 - 16\pi G c_4
R_{\alpha\beta\lambda\sigma} R^{\alpha\beta\lambda\sigma}) \,\, , \eqno
(16)$$
removes the term proportional to $c_4$ in $S_2$ from the gravitational
action.  Of course such a redefinition of the metric also effects the
equation of motion of test particles.  If test particles, moving under
the influence of gravity, travel along geodesics then after the field
redefinition in eq. (16) the equation of motion becomes
$$	{d^2 x^\nu\over ds^2} + \Gamma_{\alpha\beta}^\nu {dx^\alpha\over ds}
{}~ {dx^\beta\over ds} + 8\pi G c_4 (R_{\alpha\beta\lambda\sigma}
R^{\alpha\beta\lambda\sigma})^{;\nu} $$
$$	- 8\pi Gc_4 (R_{\alpha\beta\gamma\delta}
R^{\alpha\beta\gamma\delta})_{;\lambda} {dx^\lambda\over ds} ~
{dx^\nu\over ds} = 0\,\, . \eqno (17)$$
The temperature and mass of the black hole can be measured at infinity
and don't depend on test particles equation of motion.
Consequently the temperature mass relation is independent of $c_3$ and
$c_4$.  However, to measure the radius of the horizon a test particle
must travel near the black hole horizon where corrections to the equation
of motion (like that in eq. (17)) cannot be neglected.  That is why
$c_3$ and $c_4$ influence the relationship between the radius of the
horizon and the mass of the black hole.  We expect that typically higher
dimension operators (involving the curvature tensor) in the action for
matter fields will induce changes in the equation of motion of the
corresponding test particles like that in eq. (17).  Furthermore, these
changes will, in general, be different for different types of test
particles so that a single redefinition of the metric will not make all
test particles travel along geodesics.  In this case the value of the
radius of the horizon depends on which type of test particle is
performing the measurement.  (For related reasons such higher dimension
operators cause the flux of Hawking radiation from a black hole to depend
on the type of particle being radiated.)

Applying the thermodynamic$^{[8]}$ identity $dE = Td{\bf S}$ with $E$ equal
to the mass of the black hole and $T$ the Hawking temperature gives
$$	{d{\bf S}\over dM} = {1\over T} = 8\pi GM - {8\pi^2\over G^2
M^3} (2c_1 - {1\over 2} c_2) \,\, , \eqno (18)$$
where eq. (15) was used for $T$.  The entropy ${\bf S}$ can also be calculated
from the Euclidean path integral using the method of Gibbons and
Hawking.$^{[9]}$  The terms proportional to $c_1$ and $c_2$ in $S_2$
contribute to the Euclidean action and thus to the free energy divided
by the temperature.  The
Euclidean geometry corresponds to the analytic continuation of the part
of the Lorentzian geometry outside the horizon.  Using
$R_{\mu\nu\lambda\sigma} R^{\alpha\beta\lambda\sigma}
R^{\mu\nu}_{~~~\alpha\beta} = 12 (2GM)^3/r^9$ and $R_{~~\lambda\sigma}^{\mu\nu}
R_{\mu\alpha}^{~~~\lambda\beta} R^{\alpha \sigma}_{~~~\nu\beta} = - 3
(2GM)^3/r^9$ the contribution of the terms in $S_2$ to the free energy
divided by the temperature is
$$	\eqalign{S_2^{(Euclidean)} &= - (8\pi GM) 4\pi \left({12c_1
- 3c_2\over 6}\right) {1\over (2GM)^3}\cr
& = - {4\pi^2\over G^2M^2} \left(2 c_1 - {1\over 2} c_2\right) \,\, .} \eqno
(19)$$
The equation of motion $G^{\mu\nu} = 8\pi G {\cal T}_2^{\mu\nu}$, with the
explicit form for ${\cal T}_2^{\mu\nu}$ in eq. (11), implies (when the
indices are contracted with the metric $g_{\mu\nu}$) that the
contribution of $S_0$ to the free energy divided by the temperature is
the same as $S_2$.  Finally the surface term involving the extrinsic
curvature gives the usual contribution, $M/2T$, to the free energy
divided by the temperature.$^{[10]}$  Putting these results together
yields the entropy,
$$	\eqalign{{\bf S} &= {M\over T} - {F\over T}\cr
&= {M\over 2T} - 2S_2^{(Euclidean)}\cr
&= 4\pi GM^2 + {4\pi^2\over G^2 M^2} (2c_1 - {1\over 2} c_2) \,\, .\cr}
\eqno (20)$$
Differentiating this with respect to the black hole mass we see that the
calculation of the entropy using the Euclidean path integral method of
Gibbons and Hawking agrees with the naive thermodynamic relation in eq.
(18).  If a term proportional to $R_{\alpha\beta\lambda\sigma}
R^{\alpha\beta\lambda\sigma}$ was added to $S_1$ it would also
contribute to the Euclidean action and hence to the entropy.  However,
this is consistent with the thermodynamic relation in eq. (18) since
such a contribution of $S_1$ to the entropy is a constant independent of
$M$.  The agreement between eqs. (20) and (18) provides further evidence that
in the
semiclassical approximation the prescription of Gibbons and Hawking for
the Euclidean path integral correctly gives the free energy of the black
hole.

The Euclidean path integral and thermodynamic calculations of the
entropy agree, however, the entropy is not equal to $\pi r_H^2/G$ (i.e.,
one quarter the area of the horizon in appropriate units).  Since one-loop
Feynman diagrams
can generate operators like those in $S_2$ we expect that even if the
gravitational action were given by $S_0$  quantum corrections of order
$\hbar$ would cause a deviation from the relation ${\bf S} = \pi
r_H^2/G$.  (Quantum corrections to the entropy area relation of order $e^{-
\kappa/\hbar}$ have previously been considered$^{[11,12]}$.)

The major
conclusions of this paper, that the Euclidean path integral correctly
gives the free energy and that the entropy is not 1/4 the area of the
horizon, have already been observed in Ref. [1] for the case of gravity
coupled to
a dilaton field and the particular higher powers of the curvature tensor
implied by string theory.  (Similar conclusions have also been made for
Lovelock gravity$^{[13]}$.)  Our results demonstrate these conclusions in
the slightly different case of pure gravity (i.e., no dilaton) in four
spacetime dimensions with generic  higher dimension operators.

We thank J. Preskill, A. Strominger and S. Trivedi for helpful
discussions.

\noindent {\bf References}

\item{1.}  C.G. Callan, R.C. Myers, M.J. Perry, Nucl. Phys., {\bf B311},
673 (1988); R.C. Myers, Nucl. Phys., {\bf B289}, 701 (1987).

\item{2.}  S.W. Hawking, Commun. Math. Phys., {\bf 43}, 199 (1975).

\item{3.}  R.C. Myers and J.Z. Simon, General Relativity and
Gravitation, {\bf 21}, 761 (1989).

\item{4.}  We assume the cosmological constant vanishes.

\item{5.}  M. Visser, Phys. Rev., {\bf D46}, 2445 (1992).

\item{6.}  See for example, T. Veltman in {\it Methods in Field Theory},
Les Houches 1975, edited by R. Ballian and J. Zinn-Justin, North Holland
Publishing Co., Amsterdam (1976).  Note, however, that it is not the
divergence of a four-vector.

\item{7.}  At zeroth order (i.e., in the Schwarzschild background)
$G^{\mu\nu} = 0$.  See Ref. [5] for an explicit expression for the
components of the Einstein tensor evaluated in the metric of eq. (3).

\item{8.}  J.D. Bekenstein, Phys. Rev., {\bf D7}, 2333 (1973).

\item{9.}  G.W. Gibbons and S.W. Hawking, Phys. Rev., {\bf D15}, 2752
(1975).

\item{10.}  For a very clear discussion of this contribution see: S. Coleman,
J. Preskill and F. Wilczek, Nucl. Phys., {\bf B378}, 175 (1992).

\item{11.}   F. Dowker, R. Gregory, J. Traschen, Phys. Rev., {\bf D45},
2762 (1992).

\item{12.}  I. Moss, Phys. Rev. Lett., {\bf 69}, 1852 (1992).

\item{13.}  R.C. Myers and J.Z. Simon, Phys. Rev., {\bf D38}, 2434
(1988); D.L. Wittshire, Phys. Rev., {\bf D38}, 2445 (1988).
\bye